\documentclass[a4paper]{jpconf}
\usepackage{graphicx}
\usepackage{amsmath}

\newcommand{\op}[1]{\hat{#1}}

\renewcommand{\vec}[1]{\mathbf{#1}}

\newcommand{\ket}[1]{\ensuremath{\,|{#1}\rangle}}
\newcommand{\braket}[2]{\ensuremath{\langle{#1}|{#2}\rangle}}

\newcommand{\cg}[6]{\biggl\langle\,\begin{matrix} {#1} & {#3} \\ {#2} & {#4} \end{matrix}\, \biggr|\biggl.\, \begin{matrix} {#5} \\ {#6} \end{matrix} \,\biggr\rangle}

\newcommand{\fm}{\ensuremath{\textrm{fm}}}
\newcommand{\MeV}{\ensuremath{\textrm{MeV}}}
\newcommand{\nuc}[2]{\ensuremath{^{#1}\mathrm{#2}}}

\begin{document}

\title{The Hoyle state and its relatives}

\author{Thomas Neff$^{1}$ and Hans Feldmeier$^{1,2}$}

\address{$^1$ GSI Helmholtzzentrum f{\"u}r Schwerionenforschung GmbH and\\ ExtreMe Matter Institute EMMI,
  Planckstra{\ss}e 1, 64291 Darmstadt, Germany}
\address{$^2$ Frankfurt Institute for Advanced Studies, Max-von-Laue Stra{\ss}e 1, 60438 Frankfurt, Germany}

\ead{t.neff@gsi.de}

\begin{abstract}
The Hoyle state and other resonances in the continuum above the 3 $\alpha$ threshold in $^{12}$C are studied in a microscopic cluster model. Whereas the Hoyle state is a very sharp resonance and can be treated reasonably well in bound state approximation, the other higher lying states require a proper treatment of the continuum. The model space consists of an internal region with 3 $\alpha$ particles on a triangular grid and an external region consisting of the $^8$Be ground state and excited (pseudo)-states of $^8$Be with an additional $\alpha$. The microscopic $R$-matrix method is used to match the many-body wave function to the asymptotic Coulomb behavior of bound states, Gamow states and scattering states. $^8$Be-$\alpha$ phase shifts are analyzed and resonance properties like radii and transition strengths are investigated. 
\end{abstract}

\section{Introduction}

It is surprising that the structure of a stable nucleus like \nuc{12}{C} is still very much under discussion. The first excited $0^+$ state, the famous Hoyle state, has been in the focus of innumerable studies in recent years. This is motivated both by the exotic properties of this state, like a very large extension confirmed by precise electron scattering data \cite{hoyle07,hoyle10}, and the fact that it provides both a challenge and a benchmark for nuclear structure models. Microscopic cluster models have been able to describe many properties of the Hoyle state \cite{kamimura78,kamimura81}. More recently it was shown that the same results could be obtained with the comparatively simple THSR ansatz \cite{funaki03}. Whereas the $\alpha$ cluster structure in these calculations has been assumed from the beginning, AMD \cite{kanadaenyo07} and FMD \cite{hoyle07} approaches use a Gaussian wave-packet basis and cluster structures appear naturally in a variational procedure. The Hoyle state is also a challenge for \
textit{ab initio} methods that try to solve the many-body problem exactly for realistic two- and three-body interactions. Within the harmonic oscillator basis it is extremely difficult to describe the asymptotic behavior of three loosely interacting $\alpha$-particles \cite{fmd12}. This problem can be addressed by the symmetry-adapted NCSM \cite{dreyfuss13}. A new \emph{ab initio} approach, that is not based on a wave function method, is  using chiral perturbation theory on a lattice with Monte-Carlo techniques \cite{epelbaum11}.

The Hoyle state is located just above the three $\alpha$ threshold and has a very narrow width of only 8.5~eV. This is not the case for other cluster states that have been investigated in recent years. The properties of the second $2^+$ state could be determined unambiguously by excitation with real photons \cite{zimmerman13,weller13}. A second $4^+$ state (actually lower in energy than the $4^+$ member of the ground state band) was found in \cite{freer11}. In addition negative parity states were identified \cite{marin14}. Nevertheless there are still open questions. For example $\beta$-decays from \nuc{12}{B} and \nuc{12}{N} apparently do not populate the second $2^+$ state mentioned above but another higher lying $2^+$ state \cite{hyldegaard10}. From the theoretical point of view these states can be discussed in a meaningful way only if the continuum is treated properly. 

Our goal is to perform a microscopic calculation within fermionic molecular dynamics (FMD) \cite{fmd08} including a \nuc{8}{Be}+\nuc{4}{He} continuum in the spirit of our calculation for the $\nuc{3}{He}(\alpha,\gamma)\nuc{7}{Be}$ cross section \cite{neff11}. As a preparation for the full calculation we perform a simplified calculation within the microscopic cluster model that is presented in this contribution. Compared to earlier studies using a very similar approach \cite{descouvemont87,arai06} a larger basis is used for the \nuc{8}{Be} wave functions and a larger number of \nuc{8}{Be} pseudo-states is included in the calculation.

\section{Microscopic Cluster Model}

\begin{figure}
  \centering
  \includegraphics[width=0.5\textwidth]{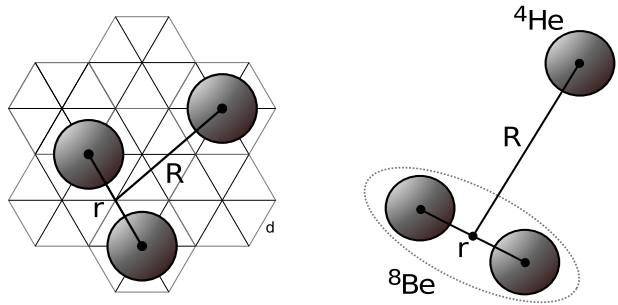}
  \caption{In the internal region three-alpha configurations are generated by putting three alphas on a triangular grid (left). \nuc{8}{Be}-$\alpha$ configurations are characterized by the distance $R$ between \nuc{8}{Be} eigenstates and the third alpha (right).}
  \label{fig:configs}
\end{figure}

We use the microscopic cluster model with Brink-type wave functions, where basis states are given as antisymmetrized intrinsic wave functions of three $\alpha$ clusters, projected on parity and angular momentum:
\begin{equation}
 \ket{\Psi^{3\alpha}_{JMK\pi}(\vec{R}_1,\vec{R}_2,\vec{R}_3)} =
      \op{P}^\pi \op{P}^J_{MK} \op{\mathcal{A}} \left\{ 
        \ket{\Psi^\alpha(\vec{R}_1)} \otimes
        \ket{\Psi^\alpha(\vec{R}_2)} \otimes 
        \ket{\Psi^\alpha(\vec{R}_3)} \right\} \: .
  \label{eq:gcm}
\end{equation}
As in our previous work \cite{hoyle07} we use the Volkov~V2 interaction and the $\alpha$ particle parameters proposed by Kamimura \cite{kamimura78,kamimura81} that provide Hoyle state properties in good agreement with experiment. In the internal region an efficient way to obtain a complete basis of three-$\alpha$ configurations is to put the three $\alpha$'s on a triangular grid as shown in Fig.~\ref{fig:configs}. The grid spacing parameter $d$ is chosen as 1.75~fm. The basis states can then be sorted according to the hyperradius $\rho^2 = \tfrac{1}{2} \vec{r}^2 + \tfrac{2}{3} \vec{R}^2$ with $\vec{r} = \vec{R}_1 - \vec{R}_2$ and $\vec{R} = \tfrac{1}{2} (\vec{R}_1 + \vec{R}_2) - \vec{R}_3$.

In the external region \nuc{8}{Be}+$\alpha$ configurations are used. The \nuc{8}{Be} eigenstates are obtained by diagonalizing $\alpha$-$\alpha$ configurations projected on angular momentum with distances up to 10~fm:
\begin{equation}
 \ket{\Psi^{\nuc{8}{Be}}_{IK}} = 
 \sum_i \op{P}^I_{K0} \op{\mathcal{A}} \left\{ \ket{\Psi^\alpha(-\tfrac{r_i}{2}\vec{e}_z} \otimes \ket{\Psi^\alpha(+\tfrac{r_i}{2}\vec{e}_z} \right\} c_i^I \: .
 \label{eq:be8}
\end{equation}
We include the $0^+$ ground state, the second $0^+$ state, two $2^+$ states and a $4^+$ state as \nuc{8}{Be} configurations. Whereas the ground state is a very narrow resonance (with the Volkov interaction used here it is actually bound by 50~keV) the $2^+$ and $4^+$ states are very wide resonances. With the exception of the ground state, the eigenstates used here therefore should be considered as pseudo-states that allow us to improve the description of the continuum. In principle the correct asymptotics would require to match to a real three-body Coulomb continuum. However this is not doable in a microscopic approach. Furthermore experiment tells us that the \nuc{12}{C} resonances with natural parity predominantly decay through the \nuc{8}{Be} ground state. 

\begin{figure}
 \centering
 \includegraphics[width=0.48\textwidth]{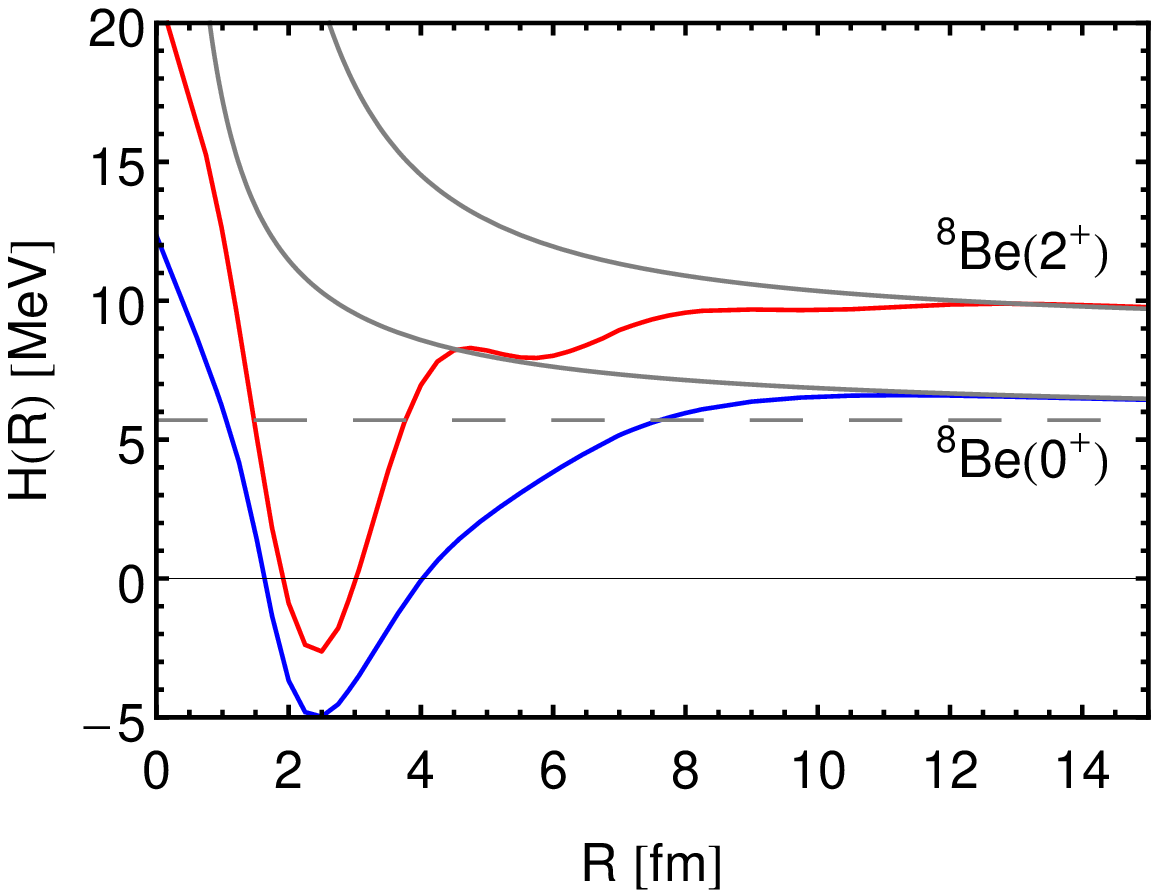}\hfil
 \includegraphics[width=0.48\textwidth]{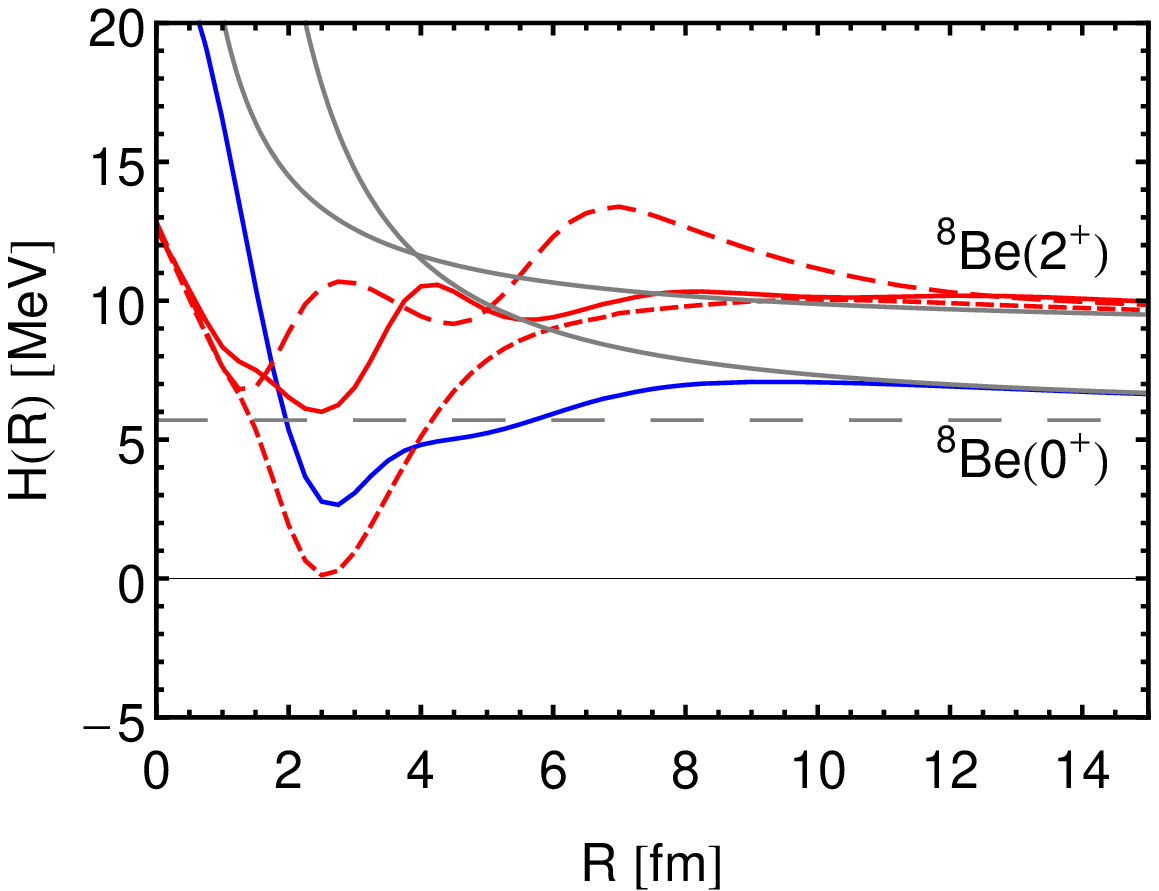}
 \caption{GCM \nuc{8}{Be}-$\alpha$ energy surfaces for total spin $0^+$ (left) and $2^+$ (right). The blue curves correspond to the cluster configurations with the \nuc{8}{Be} ground state and the red curves to the different $K$-projections of the $2^+$ state in \nuc{8}{Be}. For total spin $0^+$ only $K=0$ is possible, for total spin $2^+$ we have $K=0,1,2$. $K=2$ (red short-dashed) gives the most attraction. The gray curves denote the sum of the threshold energies, the centrifugal energy and the Coulomb energy between the clusters. The gray dashed line indicates the kinetic energy of the relative motion for the GCM basis states due to localization.}
 \label{fig:energysurface}
\end{figure}

The \nuc{8}{Be}-$\alpha$ basis states depend on the \nuc{8}{Be} eigenstate (with angular momentum $I$ and projection $K$) and the generator coordinate $R_j$ for the distance between the \nuc{8}{Be} and $\alpha$ clusters. The product wave function has to be projected on total angular momentum $J$, projection $M$ and parity $\pi$: 
\begin{equation}
 \ket{\Psi^{\nuc{8}{Be},\alpha}_{IK;JM\pi}(R_j)} = 
      \op{P}^\pi \op{P}^J_{MK} \op{\mathcal{A}} \left\{ \ket{\Psi^{\nuc{8}{Be}}_{IK}(-\tfrac{R_j}{3} \vec{e}_z)} \otimes \ket{\Psi^{\alpha}(+\tfrac{2R_j}{3} \vec{e}_z} \right\} \; .
  \label{eq:be8-alpha}
\end{equation}
Evaluating matrix elements with these basis states leads to a five-dimensional integration and therefore a large numerical effort. The GCM energy surfaces shown in Fig.~\ref{fig:energysurface} give us some first hints about the \nuc{12}{C} structure. From the minima in the energy surfaces we conclude that \nuc{8}{Be}-$\alpha$ configurations with a distance of about 2.5~fm should have a large overlap with the members of the ground state band. Furthermore both the $0^+$ ground state and the $2^+$ state in \nuc{8}{Be} appear to be important. At large distances the lowest energy surface always (for natural parity states) corresponds to the \nuc{8}{Be} ground state curve. The energy surfaces also tell us that the Coulomb barrier is located at pretty large distances of about 10-12~fm and has a height of about 1.5~MeV. This is consistent with a spatially very extended but also very sharp Hoyle state resonance, that sits just above the threshold. On the other hand we would expect rather large resonance widths for 
states significantly above the threshold, as the barrier is not very high.

In the GCM basis states the relative motion of the clusters is entangled with the total center of mass motion. To perform the matching to the asymptotic Coulomb solutions the GCM basis states have to be rewritten in terms of RGM basis states \cite{baye77}: 
\begin{equation}
  \ket{\Psi^{\nuc{8}{Be},\alpha}_{IK;JM\pi}(R_j)} = 
  \sum_L \cg{I}{K}{L}{0}{J}{K} \int dr \, r^2 \, \sqrt{\frac{2L+1}{4\pi}} \Gamma_L(R_j;r) \, 
  \ket{\Phi^{\nuc{8}{Be},\alpha}_{(IL)JM\pi}(r)} \otimes \ket{\Phi^{\mathrm{cm}}} \: ,
\end{equation}
with the Gaussians localized at $R_j$ projected on orbital angular momentum ($\mu_A=8/3$):
\begin{equation}
  \Gamma_L(R_j;r) = 4\pi \left( \frac{\mu_A}{\pi a} \right)^{3/4} \exp \left( -\mu_A \frac{r^2+R_j^2}{2a} \right) i_L \left(\mu_A \frac{r R_j}{a} \right) \: .
\end{equation}

In the RGM basis we no longer have the projection $K$ of the \nuc{8}{Be} angular momentum onto the symmetry axis but the relative orbital angular momentum $L$ as a quantum number. The overlap of RGM basis states is given by the RGM norm kernel that becomes orthogonal only at large distances $r$ between the \nuc{8}{Be} and $\alpha$ clusters:
\begin{equation}
 N_{c,c'}(r,r') = \braket{\Phi_c(r)}{\Phi_{c'}(r')} 
 \overset{r,r' \rightarrow \infty}{\longrightarrow} \delta_{cc'} \frac{\delta(r-r')}{rr'} \:
\end{equation}
We use here a short-hand notation $c = \{(IL)JM\pi \}$ for the channel quantum numbers. With the help of the RGM norm kernel we can map the many-body wave function $\ket{\Psi}$ onto the overlap function
\begin{equation}
 \psi_{c}(r) = \int dr' r'^2 \: N_{c,c'}^{-1/2}(r, r') \braket{\Phi_{c'}(r')}{\Psi} \: ,
\end{equation}
that can be interpreted as the relative wave function of the two clusters. It can be used for matching the many-body wave functions to the asymptotic behavior of two point-like clusters interacting only via Coulomb. To perform this matching we employ the microscopic $R$-matrix method developed by the Brussels group \cite{descouvemont10}. The matching is done at the channel radius $a = 16.5\,\fm$ outside the range of the nuclear interaction. For bound states the overlap function is matched to a Whittaker function
\begin{equation}
\psi_c(r) = A_c \: \frac{1}{r} W_{-\eta_c,L_c+1/2}(2 \kappa_c r), \qquad \kappa_c = \sqrt{-2 \mu (E-E_c)} \: , 
\end{equation}
where $A_c$ is the asymptotic normalization coefficient and $E - E_c$ are the energies with respect to the corresponding \nuc{8}{Be}-$\alpha$ thresholds. For resonances we use Gamow boundary conditions where the overlap function is matched to a purely outgoing Coulomb scattering solution with a complex energy $E = E_R - \tfrac{i}{2} \Gamma$: 
\begin{equation}
 \psi_c(r) = A_c \frac{1}{r} O_{L_c}(\eta_c, k_c r), \qquad k_c = \sqrt{2 \mu (E-E_c)} \: .
\end{equation}
Scattering states (incoming channel $c_0$) are matched to linear combinations of incoming and outgoing Coulomb solutions connected by the scattering matrix $S_{c,c_0}$:
\begin{equation}
 \psi_c(r) = \frac{1}{r} \left\{ \delta_{L_c,L_0} I_{L_c}(\eta_c, k_c r) - S_{c,c_0} O_{L_c}(\eta_c, k_c r) \right\}, \qquad k_c = \sqrt{2 \mu (E-E_c)} \: .
\end{equation}

\section{Scattering and Resonances}

\begin{figure}
 \centering
 \includegraphics[width=0.32\textwidth]{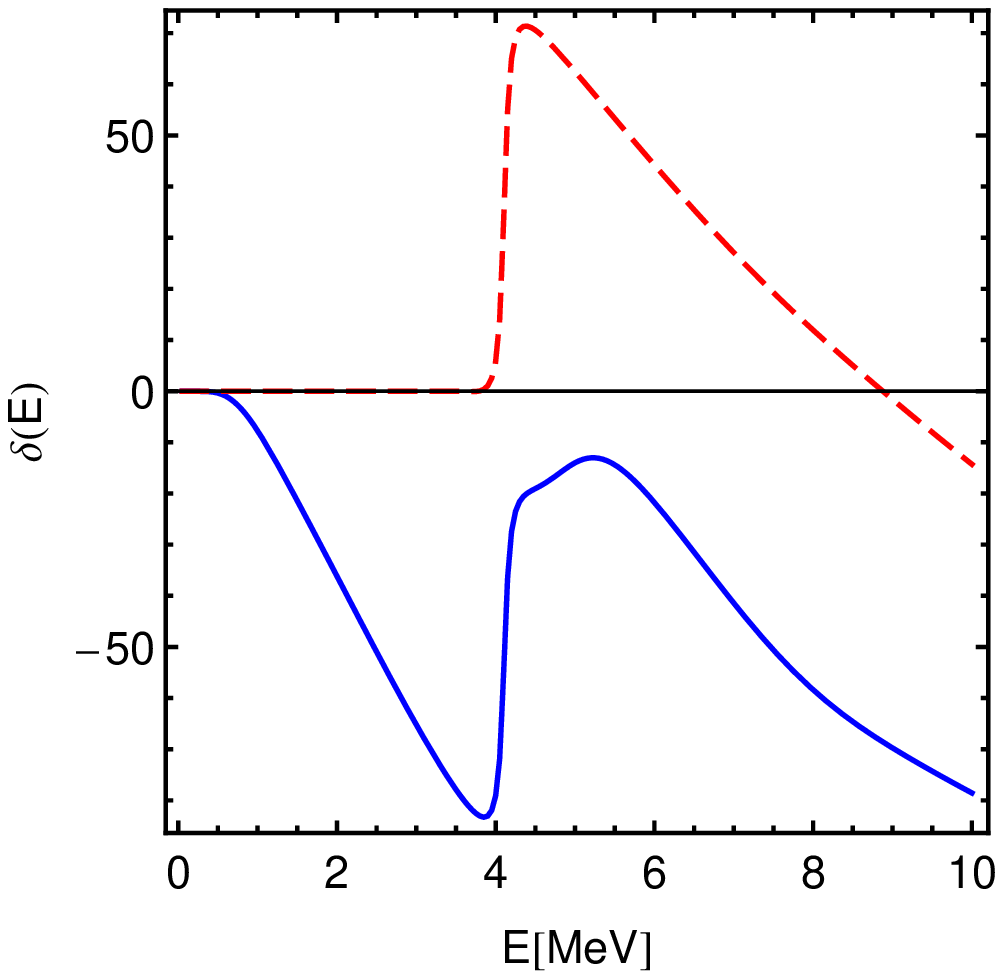}\hfil
 \includegraphics[width=0.32\textwidth]{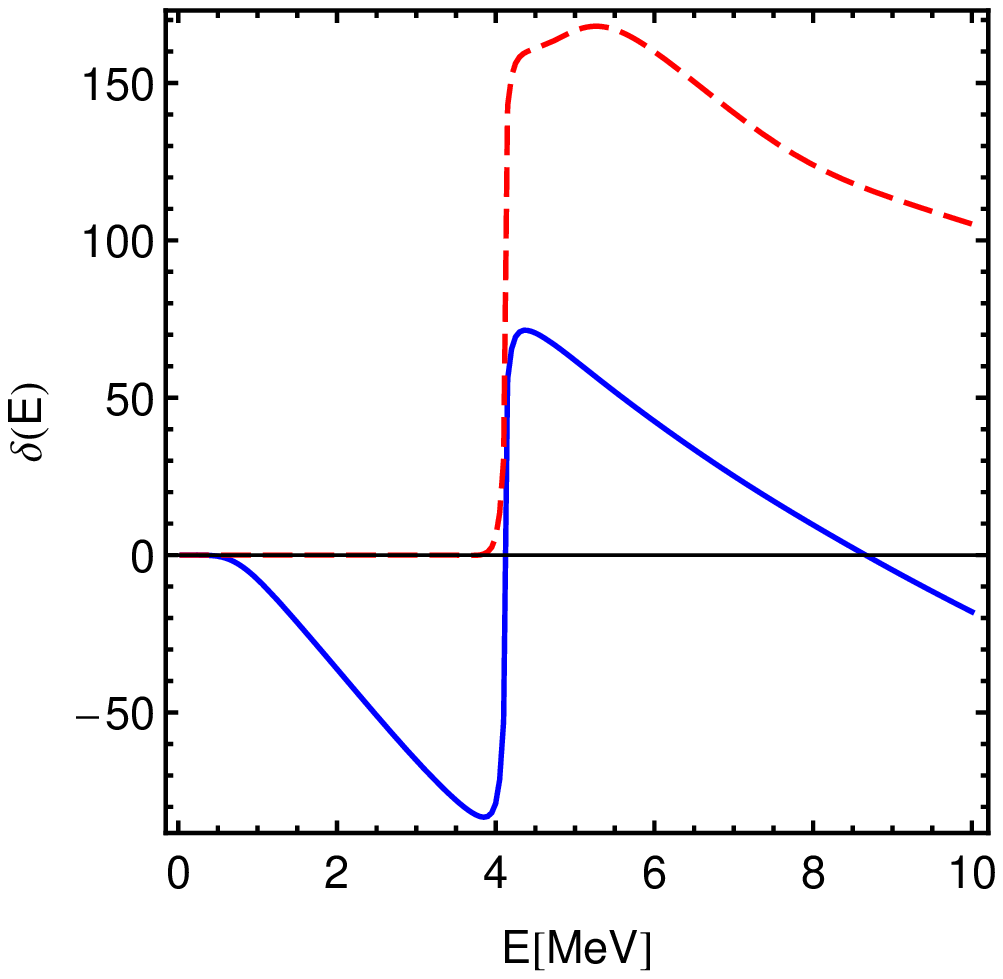}\hfil
 \includegraphics[width=0.32\textwidth]{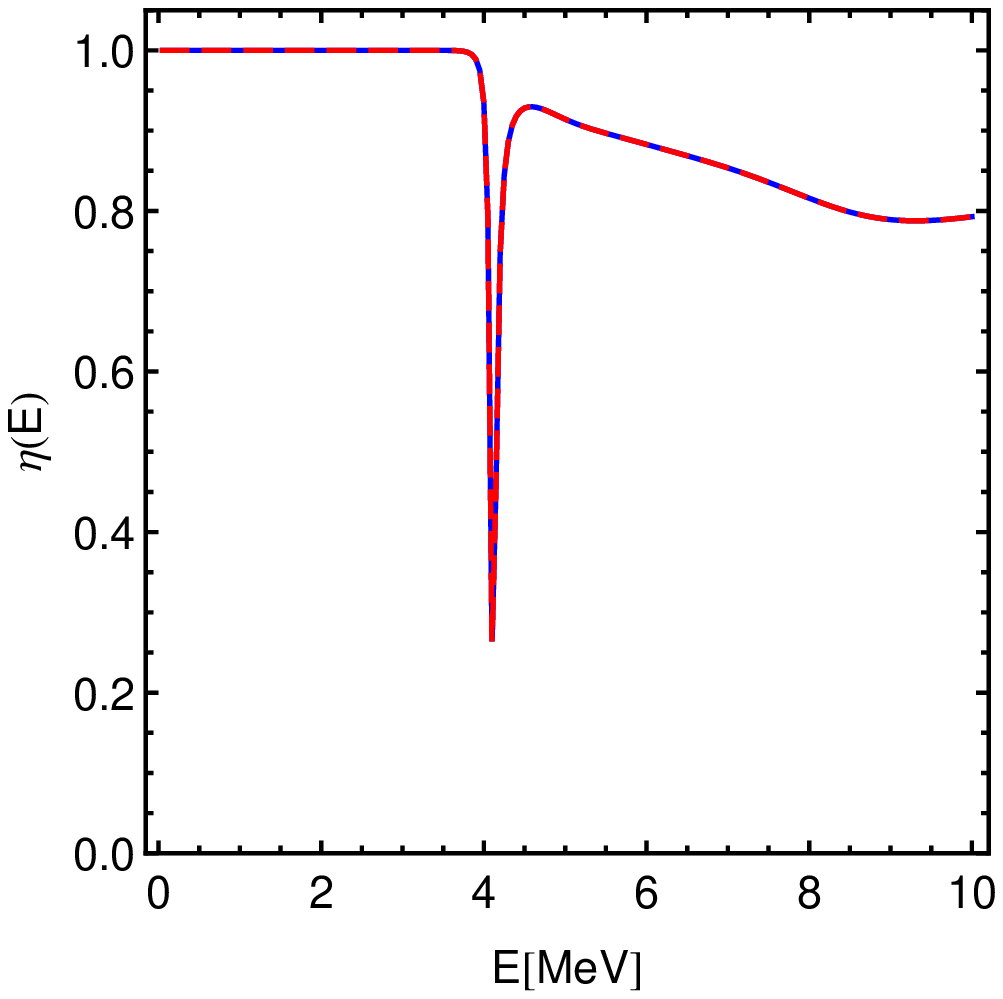}
 \caption{\nuc{8}{Be}-$\alpha$ scattering phase shifts for total spin $0^+$. The $0^+$ ground state and the first excited $2^+$ state in \nuc{8}{Be} are included in the calculation. On the left the eigenphases, in the middle the diagonal phase shifts and on the right the inelasticities. For the diagonal phase shifts the solid line (blue) indicates the \nuc{8}{Be}($0^+$) channel and dashed (red) the \nuc{8}{Be}($2^+$) channel. In case of the eigenphases the channels are mixed but have a dominant component indicated by the colors.}
 \label{fig:phaseshifts-0+}
\end{figure}

\begin{figure}
 \centering
 \includegraphics[width=0.32\textwidth]{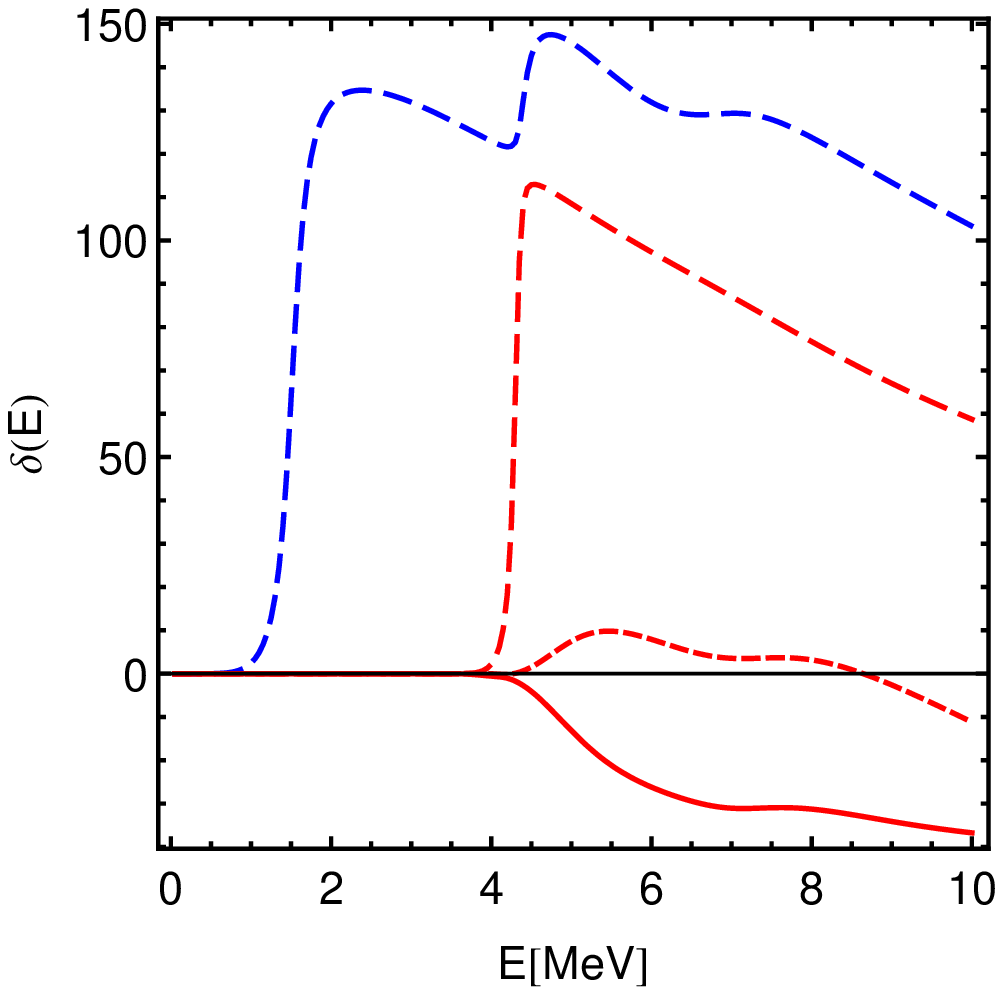}\hfil
 \includegraphics[width=0.32\textwidth]{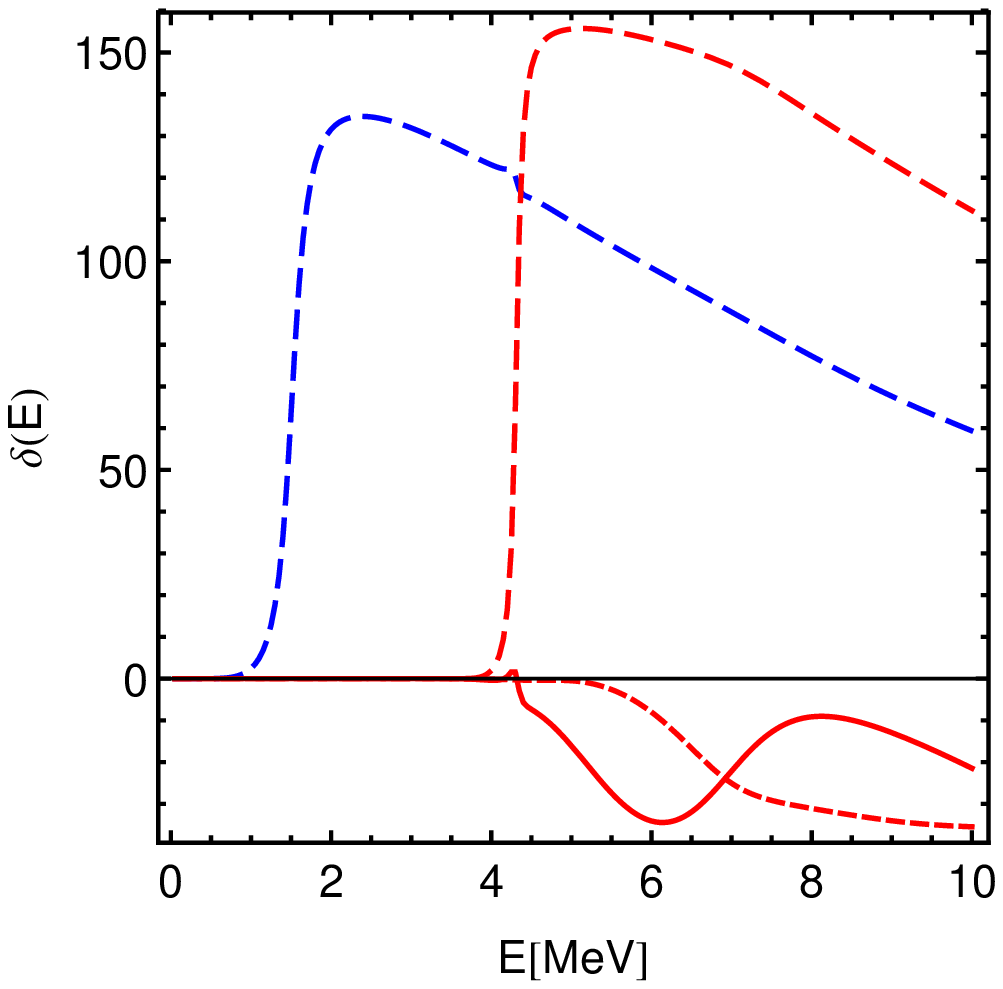}\hfil
 \includegraphics[width=0.32\textwidth]{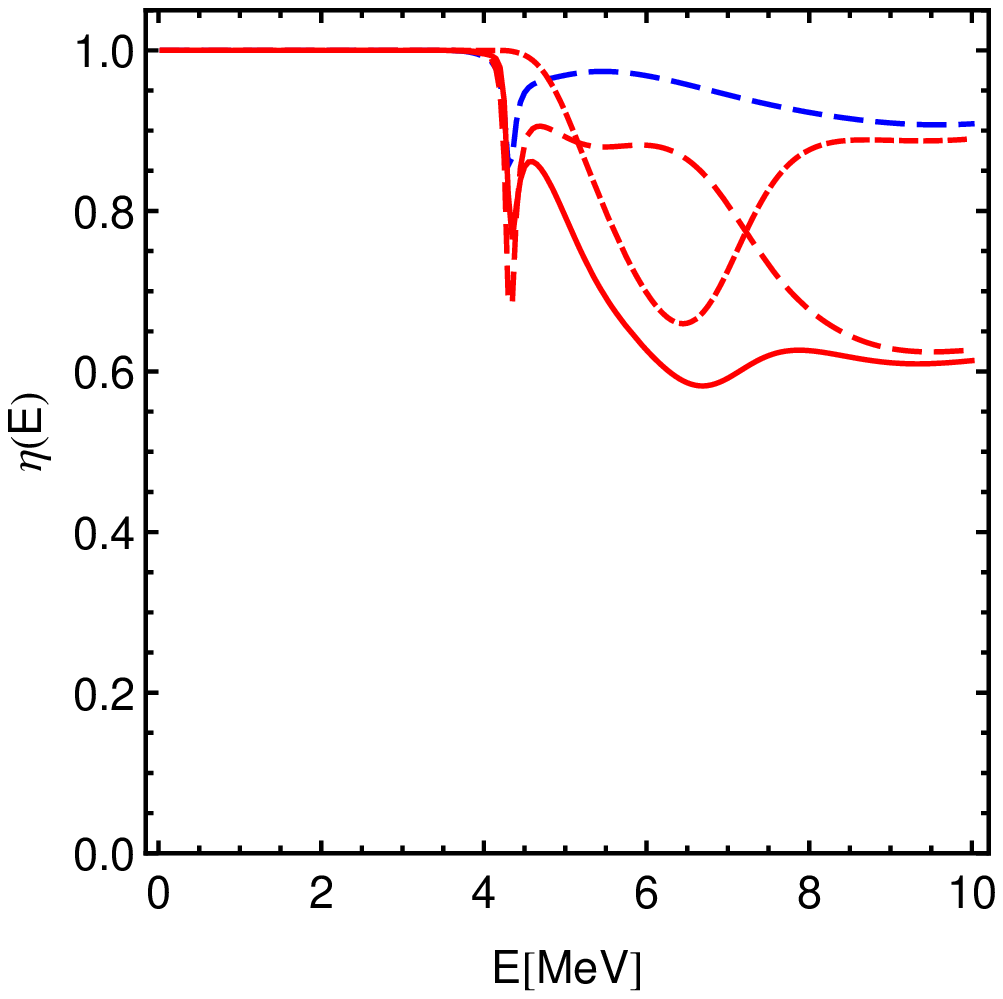}
 \caption{\nuc{8}{Be}-$\alpha$ scattering phase shifts for total spin $2^+$. There are three \nuc{8}{Be}($2^+$) channels with relative orbital angular momentum $L=0$ (solid), $L=2$ (long dashed) and $L=4$ (short dashed).}
 \label{fig:phaseshifts-2+}
\end{figure}

\begin{figure}
 \centering
 \includegraphics[width=0.32\textwidth]{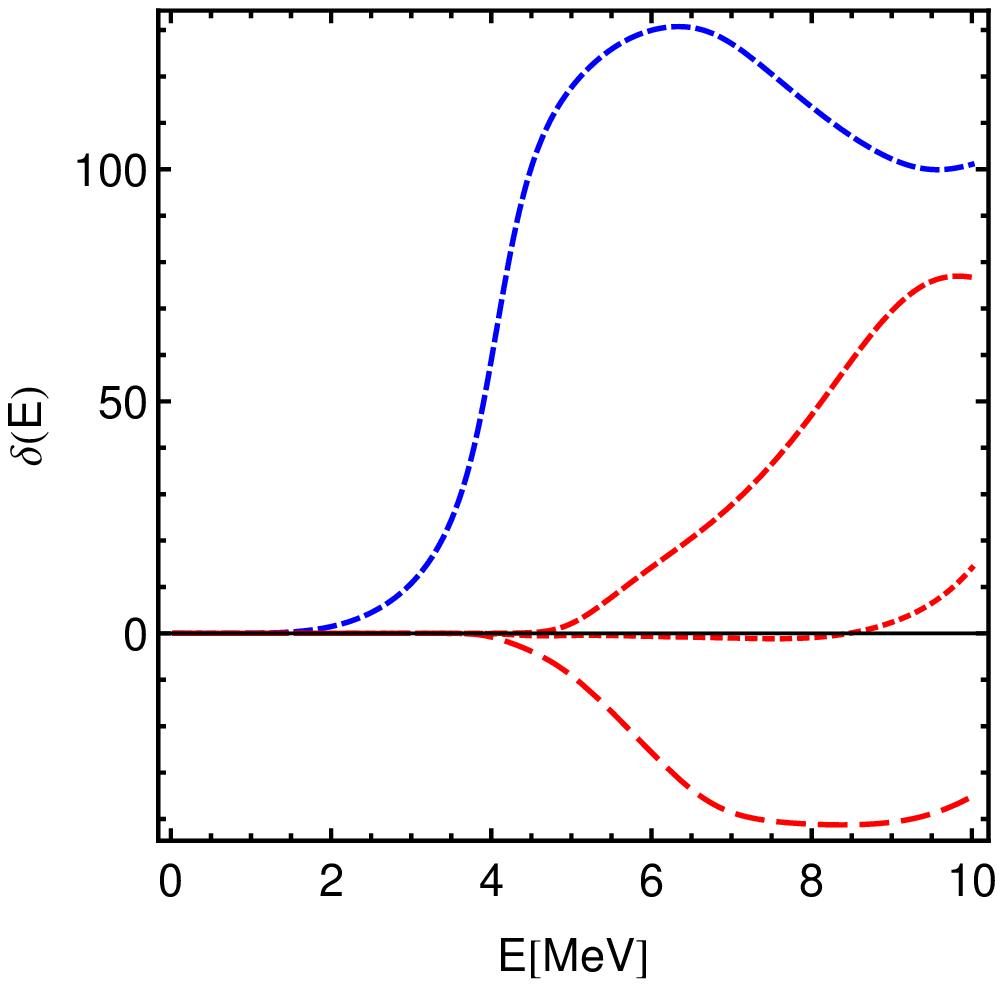}\hfil
 \includegraphics[width=0.32\textwidth]{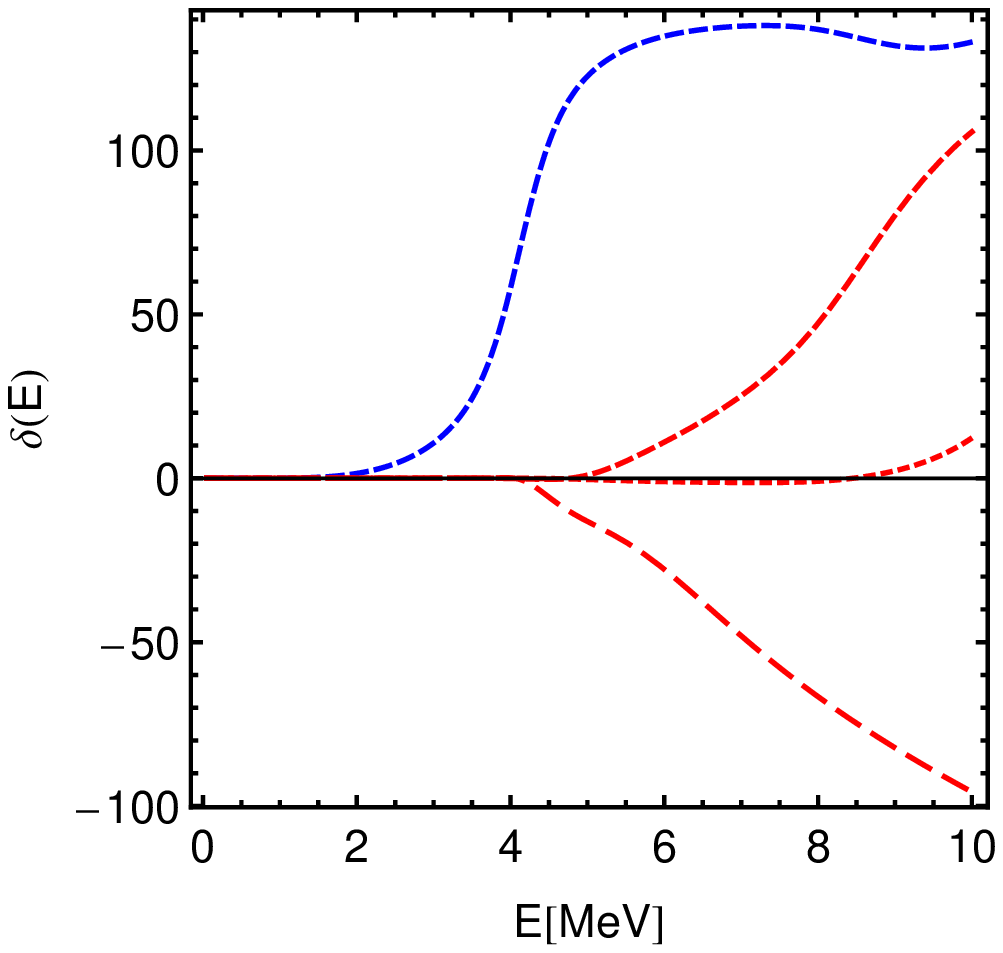}\hfil
 \includegraphics[width=0.32\textwidth]{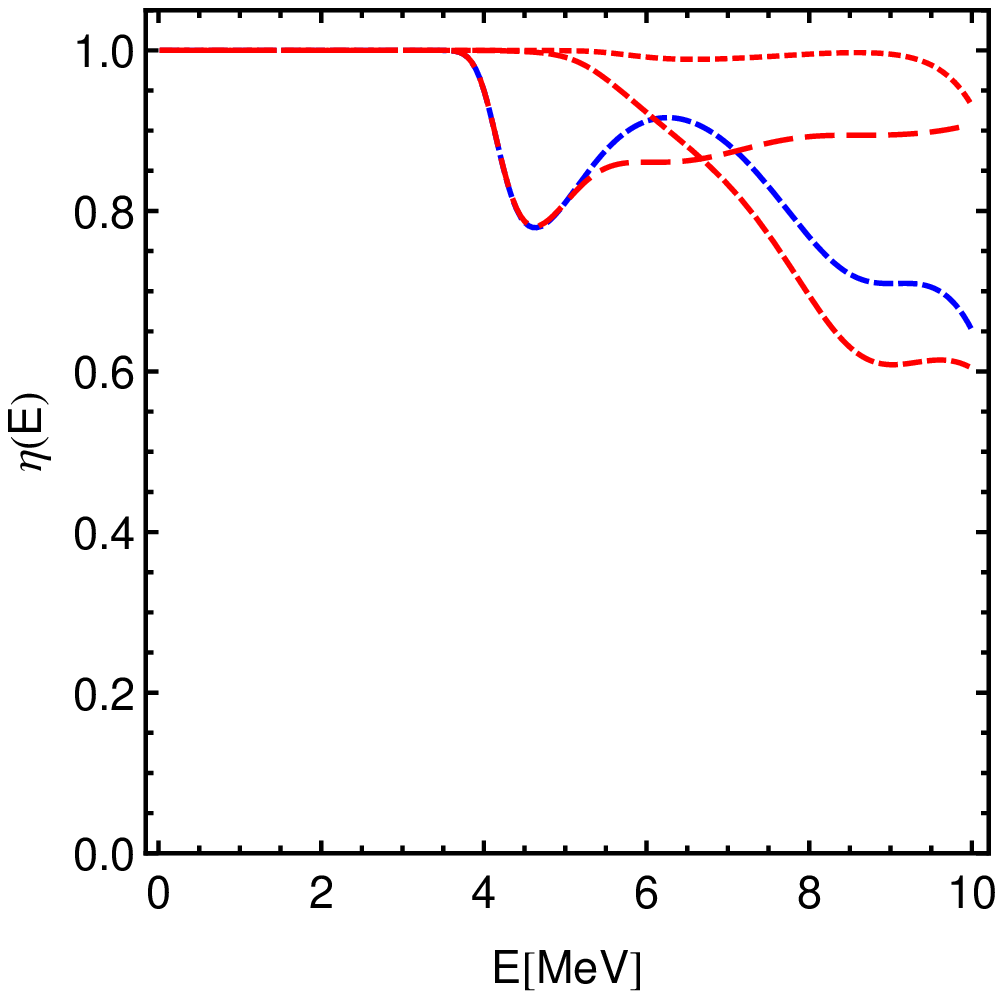}
 \caption{\nuc{8}{Be}-$\alpha$ scattering phase shifts for total spin $4^+$. There are again three \nuc{8}{Be}($2^+$) channels with relative orbital angular momentum $L=2$ (long dashed), $L=4$ (short dashed) and $L=6$ (dotted).}
 \label{fig:phaseshifts-4+}
\end{figure}

In Fig.~\ref{fig:phaseshifts-0+}-\ref{fig:phaseshifts-4+} we show the phase shifts for total spin $0^+$, $2^+$, $4^+$. The eigenphases $\delta_\alpha$ are obtained by diagonalizing the scattering matrix $S = U^{-1} D U$, $D_{\alpha\alpha} = \exp \{ 2 i \delta_\alpha \}$. The diagonal phase shifts $\delta_c$ and inelasticity parameters $\eta_c$ are obtained from the diagonal scattering matrix elements $S_{cc} = \eta_c \exp \{ 2 i \delta_c \}$. To make the plots not too busy we show here the results including only the \nuc{8}{Be} $0^+$ ground state and the first excited $2^+$ state. To calculate the phase shifts the energy is scanned in steps of 50~keV. It is therefore very difficult to resolve very narrow resonances. By employing the Gamow boundary conditions we find an additional $0^+$ resonance at an energy of 0.29~MeV and a width of 18~eV corresponding to the Hoyle state, and a $4^+$ resonance at 1.17~MeV with a width of 8~eV that is the $4^+$ member of ground state band in the cluster model. Analyzing the 
phase shifts in the $0^+$ channel we find a second resonance at 4.11~MeV and a width of 120~keV that shows a strong coupling between \nuc{8}{Be} $0^+$ and $2^+$ channels. In the $2^+$ channel we have a first resonance at 1.51~MeV ($\Gamma = 0.32\,\MeV$) that like the Hoyle state decays through the \nuc{8}{Be} ground state. There is a second resonance at 4.31~MeV ($\Gamma = 0.14\,\MeV$) that dominantly decays through the \nuc{8}{Be} $2^+$ state with relative orbital angular momentum $L=2$. In the $4^+$ channel we find besides the very narrow resonance belonging to the ground state band a resonance at 4.06~MeV ($\Gamma = 0.98\,\MeV$) that dominantly decays through the \nuc{8}{Be} ground state. 
 

The Hoyle state, the $2^+$ resonance at 1.51~MeV and the $4^+$ resonance at 4.06~MeV might be considered as members of a rotational band built on the \nuc{8}{Be} ground state with the third $\alpha$ orbiting around \nuc{8}{Be} with relative orbital angular momentum 0, 2 or 4, respectively. We also find higher lying resonances that are built on the \nuc{8}{Be} $2^+$ state. However the situation gets more complicated when we include channels based on additional \nuc{8}{Be} pseudostates. For energies below about 4~MeV we do not see significant effects on the phase shifts. At higher energies, where we find many overlapping resonances, the situation becomes even more complex. This probably indicates that we are dealing here with real three-body decays.

\section{Observables}

\begin{table}
\centering
\caption{Observables calculated for increasing model space sizes in bound state approximation and with Gamow states. Energies are given in MeV, radii in fm, $M(E0)$ matrix elements in $e\,\fm^2$, $B(E2)$ strengths in $e^2\,\fm^4$.}
\label{tab:observables}
\begin{tabular}{lccccc}
\br
                              & \shortstack{$\rho < 6\,\mathrm{fm}$\\$R<9\,\mathrm{fm}$} & \shortstack{$\rho < 6\,\mathrm{fm}$\\$R<12\,\mathrm{fm}$} & \shortstack{$\rho < 6\,\mathrm{fm}$\\$R<15\,\mathrm{fm}$} & \shortstack{$\rho < 6\,\mathrm{fm}$\\Gamow} & Experiment\\
\mr
$E(0_1^+)$                     & -89.64 & -89.64 & -89.64 & -89.64 & -92.16 \\
$E^*(2_1^+)$                   &   2.54 &   2.54 &   2.54 &   2.54 &   4.44 \\
$E^*(0_2^+)$, $\Gamma_\alpha(0_2^+)$                   
                              &   7.82 &   7.78 &   7.76 &   7.76, $3.04 \cdot 10^{-3}$ &   7.65, $8.5(10)\cdot10^{-6}$\\
$E^*(2_2^+)$, $\Gamma_\alpha(2_2^+)$
                              &   9.18 &   9.08 &   8.93 &   8.98, 0.46 & 10.13(5), $2.08^{+0.33}_{-0.26}$ \\[1ex]

$r_\mathrm{ch}(0_1^+)$       & 2.53 & 2.53 & 2.53 & 2.53 & 2.47(2) \\
$r(0_1^+)$                     & 2.39 & 2.39 & 2.39 & 2.39 & -- \\
$r(0_2^+)$                     & 3.68 & 3.78 & 3.89 & 4.08 + 0.07i & -- \\[1ex]

$B(E2, 2_1^+\!\rightarrow 0_1^+$) & 9.12 & 9.08 & 9.08 & 9.08 & 7.6(4) \\
$M(E0, 0_1^+\!\rightarrow 0_2^+$) & 6.55 & 6.40 & 6.27 & 6.02 + 0.01i & 5.47(9) \\
$B(E2, 2_2^+\!\rightarrow 0_1^+$) & 2.48 & 2.09 & 1.33 & 2.11 + 1.41i & $1.57^{+0.14}_{-0.11}$ \\
\br
\end{tabular}
\end{table}

Electromagnetic transitions are powerful tools to test the calculated wave functions. In case of \nuc{12}{C} the transition density from the ground state to the Hoyle state could be extracted from electron scattering data very precisely and confirmed the theoretical prediction of a very large radius for the Hoyle state \cite{hoyle07,hoyle10}. The theoretical calculations of the transition form factor or the transition density always employed the bound state approximation. This was assumed to be justified, as the Hoyle state is a very sharp resonance. We can now test this assumption using the Gamow states. In Tab.~\ref{tab:observables} we show the results for various observables for four model spaces. In all four model spaces we include three $\alpha$ configurations on the triangular grid up to a hyperradius $\rho$ of 6~fm. In addition we include \nuc{8}{Be}-$\alpha$ configurations with distances $R$ up to 9, 12, and 15~fm without enforcing boundary conditions. These are therefore calculations in a bound 
state approximation with increasing model space sizes. We finally show results where we use the resonance wave functions obtained as Gamow states. For the evaluation of observables we follow the procedure of Berggren \cite{berggren68,berggren96}. As expected the results for the binding energy and the radius of the ground state are the same in all model spaces. For the Hoyle state energy we can observe a small change from an excitation energy of 7.82~MeV in the smallest model space to 7.76~MeV in the largest model space. The bound state approximation is also in agreement with the resonance energy obtained for the Gamow state. Including all the \nuc{8}{Be} channels we find a resonance width of 3~keV for the Hoyle state. However, a much larger model space dependence is found for the radius of the Hoyle state. It increases from 3.68~fm to 3.89~fm in the bound state approximations and for the Gamow state we obtain a real part of the radius of 4.08~fm and a small imaginary part. The results in bound state 
approximation are consistent with our previous results \cite{hoyle07} ($r(0_2^+) = 3.71\,\fm$) where we used a different set of three $\alpha$ configurations and with the THSR results \cite{funaki03} ($r(0_2^+) = 3.83\,\fm$). This model space dependence is also seen in the monopole matrix element. The result with the Gamow state is now closer to the experimental value extracted from the electron scattering data \cite{hoyle10}. An even stronger dependence on the model space can be observed for the second $2^+$ state. The resonance width is much larger and even the resonance energy can not be determined reliably in bound state approximation. For the $B(E2)$ transition strength the results in bound state approximation differ by a factor of two. The Gamow state result is in reasonable agreement with the reanalyzed result of Zimmerman \textit{et al.} \cite{zimmerman13,weller13}. The large imaginary part indicates that one has to be careful in disentangling the resonance contribution from background contributions 
in the experimental data.

\section{Summary and Outlook}

We have presented an investigation of the \nuc{12}{C} continuum above the three $\alpha$ threshold within a microscopic $\alpha$ cluster model. In addition to the Hoyle state we find a second $2^+$ and a second $4^+$ state that are all built on \nuc{8}{Be}-$\alpha$ cluster configurations with \nuc{8}{Be} in the ground state. Resonances are described as Gamow states that can be used to calculate electromagnetic transition strengths. Alternatively we could use the scattering states to obtain strength distributions. This might be the better way to compare with experimental data, especially in case of overlapping resonances or significant background contributions. In the future we want to perform a fully microscopic calculation including continuum within the FMD approach with a realistic nucleon-nucleon interaction. This will hopefully provide us with a consistent picture of the \nuc{12}{C} structure, including for example also $\beta$-transitions and information about $T$=1 states, that can not be described in 
an $\alpha$-cluster model.

\section*{References}
\bibliographystyle{iopart-num}
\bibliography{all}

\end{document}